\documentclass[useAMS,usegraphicx]{mn2e}

\usepackage{times}
\usepackage{lscape}
\usepackage{subfigure}
\input{psfig.sty}

\def\gs{\mathrel{\raise0.35ex\hbox{$\scriptstyle >$}\kern-0.6em
\lower0.40ex\hbox{{$\scriptstyle \sim$}}}}
\def\ls{\mathrel{\raise0.35ex\hbox{$\scriptstyle <$}\kern-0.6em
\lower0.40ex\hbox{{$\scriptstyle \sim$}}}}

\makeatother

\title[The effect of FIR lines on the selection of high-$z$ galaxies]{The potential influence of far-infrared
emission lines on the selection of high-redshift galaxies}

\author[Smail et al.]{Ian Smail,$^{\! 1}$\thanks{ian.smail\@durham.ac.uk} A.\,M.\ Swinbank,$^{\! 1}$ 
R.\,J. Ivison$^{2,3}$ \& E.\ Ibar$^{3}$\\\\
$^{1}$Institute for Computational Cosmology, University of Durham, South Road,
        Durham DH1 3LE UK\\
$^{2}$UK Astronomy Technology Centre, Royal Observatory,
Blackford Hill, Edinburgh EH9 3HJ\\
$^{3}$Institute for Astronomy, University of Edinburgh,
Blackford Hill, Edinburgh EH9 3HJ\\
}

\begin{document}

\pagerange{1--6} \pubyear{2011}
\volume{666}

\maketitle 

\begin{abstract}
We investigate whether strong molecular and atomic emission lines at far-infrared wavelengths can influence the
identification and derived properties of galaxies selected from broad-band, far-infrared or submillimetre
observations.    Several of these lines, e.g.\ [C{\sc ii}]\,158$\mu$m, have been found to be very bright in some
high-redshift galaxies, with fluxes of $\gs $\,0.1--1\% of the total far-infrared luminosity, and may be even
brighter in certain populations at high redshifts.  At  redshifts where these  lines fall in instrument pass-bands
they can significantly increase the broad-band flux measurements. We estimate that the contributions from  line
emission could boost the apparent broad-band flux by $\gs $\,20--40\% in the {\it Herschel} and SCUBA-2 bands. 
Combined with the steep source counts in the submillimetre and far-infrared bands, line contamination has potentially
significant consequences for the properties of sources detected in flux-limited continuum surveys, biasing the
derived redshift distributions and bolometric luminosities.  Indeed, it is possible that some $z>4$ sources found in
850-$\mu$m surveys are being identified in part due to line contamination from strong [C{\sc ii}] emission.  These
biases may be even stronger for less-luminous and lower-metallicity populations at high redshifts  which are
observable with ALMA and which may have even stronger line-to-continuum ratios.
\end{abstract}


\section{Introduction}

Studies of the far-infrared (FIR) and submillimetre emission of our own galaxy and external galaxies identify strong
continuum emission from dust grains at a range of temperatures.  This dust emission is a key tracer of obscured
star-formation and AGN activity in the local Universe.  For high-redshift sources, the far-infrared emission is
redshifted into the submillimetre wavebands where it provides a similar census of their dust-obscured activity.  The
development of efficient far-infrared and submillimetre imagers (e.g.\ SPIRE on the {\it Herschel} Space Observatory,
Griffin et al.\ 2010; LABOCA on APEX, Siringo et al.\ 2009; or  SCUBA-2 on the JCMT, Holland et al.\ 2006) are
beginning to provide large far-infrared- and submillimetre-selected samples of luminous, high-redshift galaxies:
submillimetre galaxies (SMGs).  These SMGs represent a population of extremely luminous, but highly obscured,
starbursts at $z\gs $\,1--5, suggesting that obscured star formation is increasingly important in galaxies at higher
redshifts (e.g.\  Chapman et al.\ 2005; Dole et al.\ 2006; Wardlow et al.\ 2011).

While the bulk of the luminosity of SMGs is contributed by their dust continuum emission, the gas phase in these
galaxies also cools through strong emission lines arising from atomic and molecular transitions.   The strongest of
these are identified with atomic fine-structure transitions of Carbon, Nitrogen and Oxygen and a number of molecules
(e.g.\ CO, HCN).  These lines represent the main cooling routes for dense gas and hence they help regulate the
star-formation processes.  The most commonly studied of these lines are the rotational transitions of the CO
molecule, which are  used as a  tracer of the total H$_2$ gas content of galaxies.  However, there are several atomic
lines which are far brighter than the CO lines, e.g.\ [C{\sc ii}]\,158$\mu$m or [O{\sc i}]\,63$\mu$m.   The  [C{\sc
ii}] line  traces the cool interstellar medium and the surfaces of photo-dissociation regions, with [O{\sc i}] 
arising from warmer and denser regions.  Photo-dissociation Region (PDR) theory explains these strong lines as a
result of the intense far-UV fields photo-dissociating CO resulting in bright [C{\sc ii}] (and [O{\sc i}]) emission
(Hollenbach \& Tielens 1997).  As a result the  [C{\sc ii}] line is viewed a promising probe of the interstellar
medium in distant galaxies, providing an indicator of the H$_2$ content and extent.   Indeed, studies of local
galaxies have shown that  up to $\sim 1$\% of their  bolometric output can be emitted in the [C{\sc ii}]\,158$\mu$m
or [O{\sc i}]\,63$\mu$m lines (e.g.\ Stacey et al.\ 1991; Brauher et al.\ 2008; Fig.~1).   Even stronger line
contributions, $\sim $\,3--4\% (Madden et al.\ 1997; Rubin et al.\ 2009), are seen within low metallicity,
star-forming galaxies, suggesting that these lines may be even brighter in young galaxies at high redshifts.

Early searches for [C{\sc ii}] at high redshifts  by necessity focused on the highest-redshift far-infrared sources,
$z>4$, where the [C{\sc ii}] line is shifted into the atmospheric windows at longer wavelengths in the submillimetre
(e.g.\ Ivison et al.\ 1998; Maiolino et al.\ 2005, 2009; Bolatto et al.\ 2008;  Wagg et al.\ 2010).  Most of these
sources are powerful AGN (as well as being Ultraluminous Infrared Galaxies, ULIRGs, with L$_{\rm FIR}\geq
10^{12}$\,L$_\odot$, where L$_{\rm FIR}$ is the 8--1000-$\mu$m luminosity) and it was found that their [C{\sc
ii}]\,158$\mu$m lines were weak,  relative to L$_{\rm FIR}$. This seemed to demonstrate the same behaviour as seen in
the local Universe  (Fig.~1), where the [C{\sc ii}] emission is weak in ULIRGs and in AGN-dominated systems. A number
of possible reasons have been proposed for this, including an enhanced contribution from continuum emission in dusty,
high-ionisation regions (e.g.\ Luhman et al.\ 1998, 2003).   However, more recent observations of [C{\sc ii}] in
star-formation dominated galaxies at high redshifts (Hailey-Dunsheath et al.\ 2010; Ivison et al.\ 2010a; Stacey et
al.\ 2010; Valtchanov et al.\ 2011), all of which are ULIRGs, has shown that their [C{\sc ii}] emission is as bright
as lower-luminosity galaxies at low redshifts,  L$_{\rm [CII]}/$\,L$_{\rm FIR}\sim $\,0.1--1\% (Fig.~1).  This has
been interpreted as due to the more widely distributed star-formation activity within these systems, in contrast to
the compact nuclear starbursts seen in low-redshift ULIRGs (e.g.\ Ivison et al.\ 2010a).  The strength of [C{\sc ii}]
and other atomic lines underlines their potential for studying the star formation and the physical properties of the
interstellar medium in high-redshift galaxies.  

Such strong emission lines may make significant contributions to the broad-band fluxes of these sources at  redshifts
where  the lines fall in the relevant filters.  Indeed, Seaquist et al.\ (2004) show this effect in  local FIR-bright
galaxies, where  CO(3--2) emission typically contributes 25\% of the 850-$\mu$m  flux density. Equally,  strong
atomic lines may  influence the detectability of high-redshift SMGs, as well as less-obscured galaxies, in broad-band
imaging surveys. Emission lines would similarly bias the properties derived from the broad-band spectral energy
distributions (SED) of these systems.  In this letter we discuss the likely influence of far-infrared emission lines
on the visibility of high-redshift galaxies, including SMGs, using recent constraints on their strength.  We assume a
cosmology with $\Omega_m=0.27$, $\Omega_\Lambda=0.73$ and $H_o=71$\,km\,s$^{-1}$\,Mpc$^{-1}$.

\section{Observations of FIR lines at high redshifts}

To illustrate the potential influence of far-infrared and submillimetre emission lines on the broad-band fluxes of
high-redshift galaxies we have chosen to use a well-studied example for which there are good constraints on the
continuum and line properties: SMM\,J2135$-$0102 (Swinbank et al.\ 2010).  This is a  typical $z\sim 2$ SMG  with an
intrinsic 850-$\mu$m flux of 3\,mJy and a star formation rate of $400\pm 20$\,M$_\odot$\,yr$^{-1}$ (Ivison et al.\
2010a). Importantly, it is serendipitously positioned behind a foreground, massive cluster which gravitationally
amplifies the source by a factor of $32\times$.  As a result  it has  apparent fluxes of 0.1--0.5\,Jy in the
far-infrared and submillimetre  (Ivison et al.\ 2010a). The  brightness of this source enables a detailed study of
its continuum and, critically for our analysis, emission-line properties in the far-infrared and sub/millimetre
wavebands (Ivison et al.\ 2010a; Danielson et al.\ 2011).

We include in our  SED for   SMM\,J2135$-$0102 the main emission lines observed by Ivison et al.\ (2010a) and
Danielson et al.\ (2011): [N{\sc ii}]\,122$\mu$m, [O{\sc i}]\,145$\mu$m, and [C{\sc ii}]\,158$\mu$m, as well as the
full $^{12}$CO spectral line ladder and the two [C{\sc i}] lines.   To these we add several other lines at shorter,
currently unobserved, wavelengths, but which are known to be bright in local LIRGs and ULIRGs. To do this we select
the brightest far-infrared lines from the {\it ISO} survey of local galaxies by  Brauher et al.\ (2008): [O{\sc
iii}]\,52$\mu$m, [N{\sc iii}]\,57$\mu$m, [O{\sc i}]\,63$\mu$m and [O{\sc iii}]\,88$\mu$m.    The fluxes of these
lines are  derived from the predictions of a PDR model constrained by the observed line properties and the existing
limits on line fluxes from Ivison et al.\ (2010a).  This  PDR model is described in more detail in Danielson et al.\
(2011).  

%
%
\centerline{\psfig{file="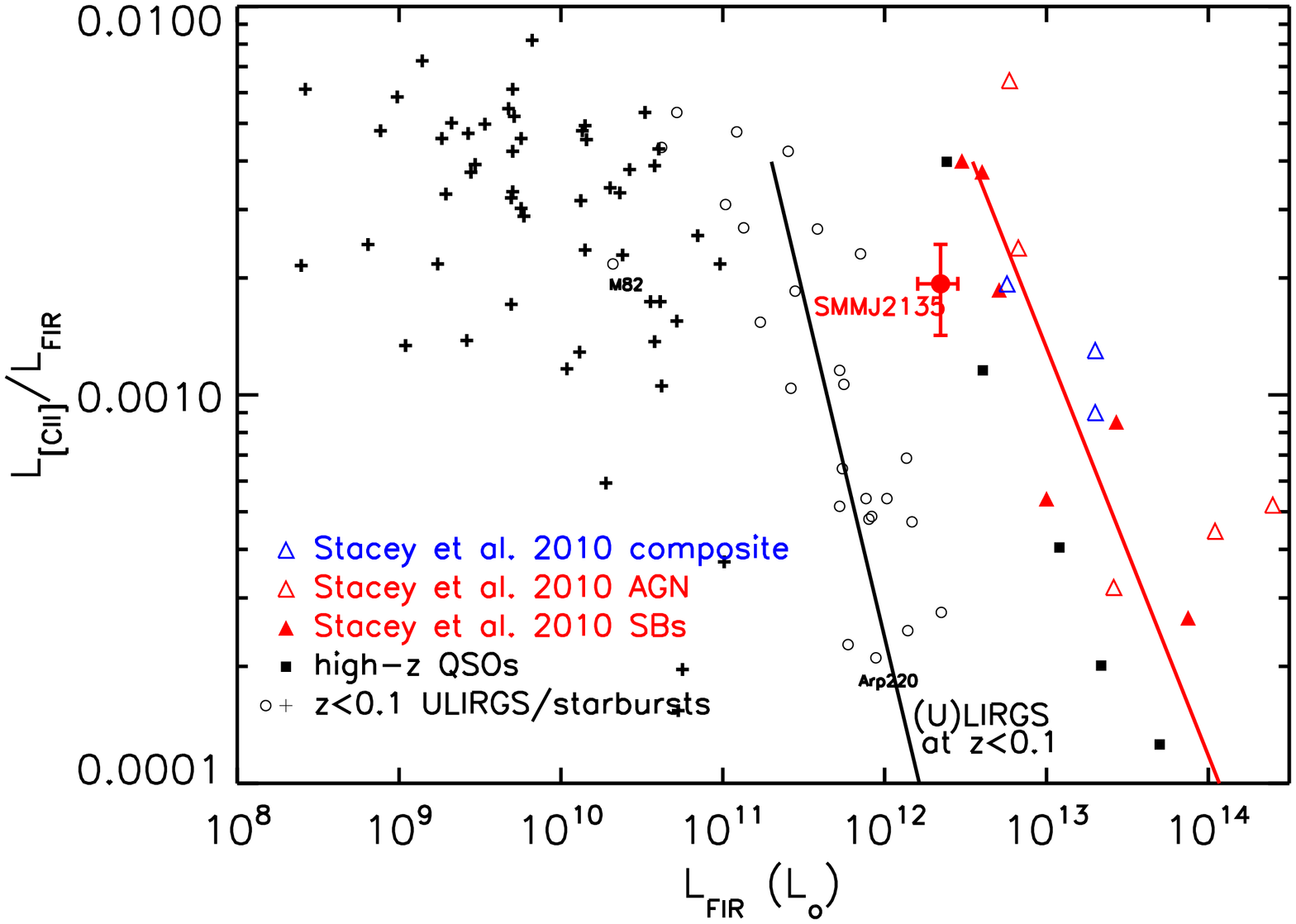",width=3.3in,angle=90}}
\noindent{\small\addtolength{\baselineskip}{-1.0pt}  {\sc Fig.~1} --- The variation in the ratio of [C{\sc
ii}]\,158$\mu$m to far-infrared luminosity for samples of low- and high-redshift starbursts, ULIRGs and AGN.   The
low-redshift samples show a steep decline in L$_{\rm [CII]}/$\,L$_{\rm FIR}$  around L$_{\rm FIR}\sim
10^{12}$\,L$_\odot$.  The earliest studies of high-redshift AGN showed similarly suppressed  [C{\sc ii}] emission. 
However, recent work on high-redshift SMGs shows much stronger  [C{\sc ii}] emission (e.g.\ Hailey-Dunsheath et al.\
2010; Stacey et al.\ 2010), including SMM\,J2135$-$0135 (Ivison et al.\ 2010a).  Indeed, even though we have only
just started to investigate the emission-line properties of SMGs, examples have been found which have  L$_{\rm
[CII]}/$\,L$_{\rm FIR}$  approaching the highest ratios known locally.  The strength of the lines in these sources
may result in significant biases towards their detection in broad-band photometric surveys. This plot is adapted from
Maiolino et al.\ (2009).

}
\smallskip

The brightest lines in this source are the observed [C{\sc ii}]\,158$\mu$m line, with a L$_{\rm [CII]}/$\,L$_{\rm
FIR}$ fraction of 0.27\%, and the predicted strength of [O{\sc iii}]\,52$\mu$m, with a L$_{\rm [OIII]}/$\,L$_{\rm
FIR}=0.37$\%, while the fractions for   [N{\sc iii}]\,57$\mu$m, [O{\sc i}]\,63$\mu$m, [O{\sc iii}]\,88$\mu$m, [N{\sc
ii}]\,122$\mu$m, [O{\sc i}]\,145$\mu$m  are 0.16\%, 0.16\%, 0.21\%, 0.19\% and 0.13\% respectively.  In contrast the
integrated $^{12}$CO ladder has a cumulative  L$_{\rm CO}/$\,L$_{\rm FIR}\sim 0.01$\%, with the two [C{\sc i}] lines
contributing just 0.003\% (Danielson et al.\ 2011).  We note that for   [O{\sc i}]\,63$\mu$m  we have reduced the
PDR-predicted flux by a factor of 80\%  to reflect the likelihood that the line may be self absorbed (Vasta et al.\
2010). This ensures that the predicted contribution is conservative. 

We plot the combined continuum and emission-line spectrum of SMM\,J2135$-$0102 in Fig.~2.  We also show the
transmission curves of the far-infrared and submillimetre pass-bands for the {\it Herschel} PACS and SPIRE and the
SCUBA-2 and LABOCA cameras.  Due to the  line-to-continuum ratios of the lines it is clear that the strongest
emission lines may have a detectable effect on the apparent brightness (and hence derived luminosity) of a source.

Before continuing, we stress that SMM\,J2135$-$0102 was first identified at 870$\mu$m, a pass-band which is not
influenced by strong line emission in this source and hence the strength of its emission lines are likely to be
representative of typical high-redshift SMGs, rather than the most extreme examples.  This is illustrated in Fig.~1,
which shows a L$_{\rm [CII]}/$\,L$_{\rm FIR}$ ratio which is not extreme for luminous, high-redshift sources, where
line emitters have already been found with $\gs 4 \times$ higher ratios.  In fact, as we discuss below, selection of
the brightest sources in a particular waveband may select a population with line-to-continuum ratios in that waveband
which are far higher than typical.
\medskip

%
%
\centerline{\psfig{file="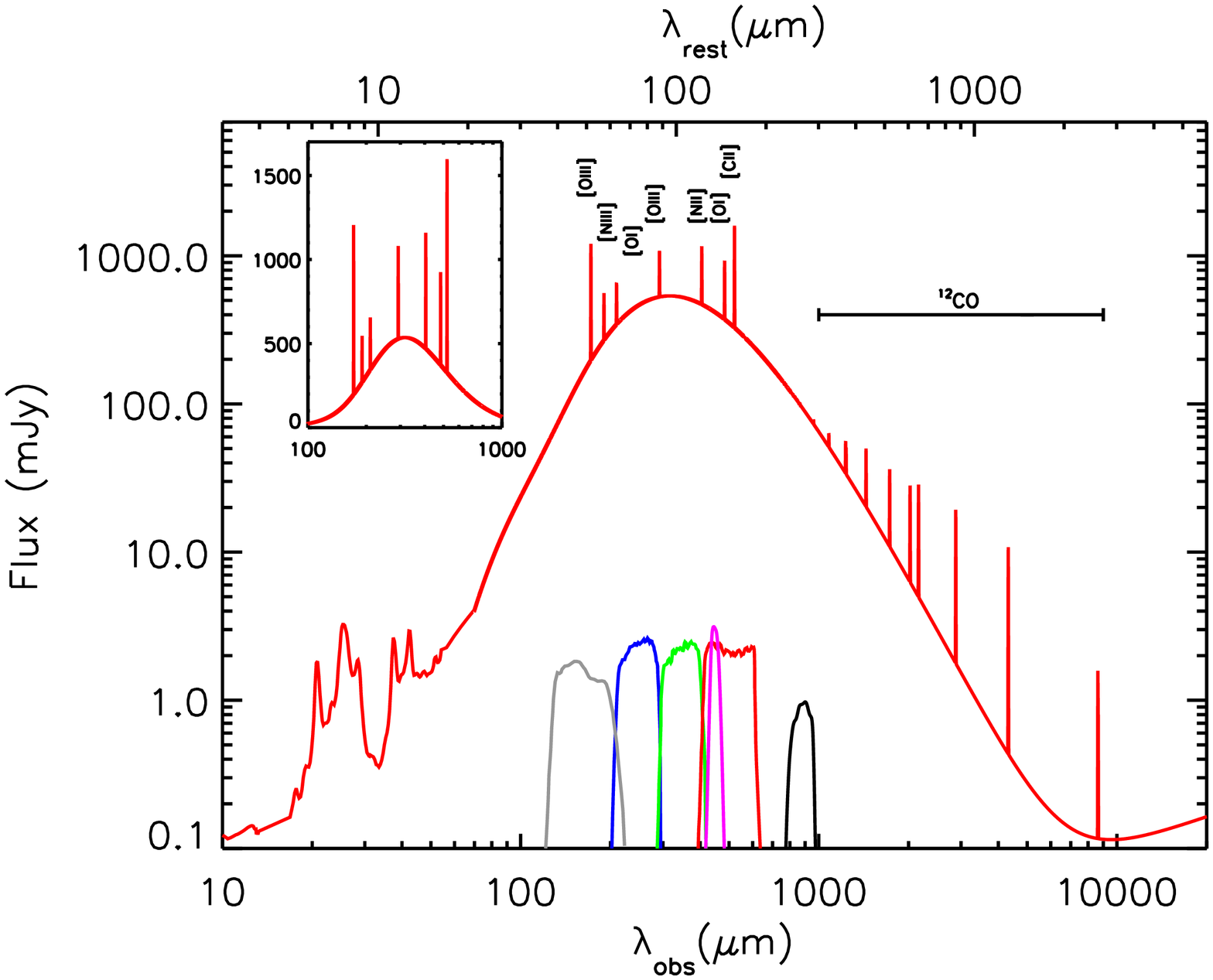",width=3.3in,angle=90}}
\noindent{\small\addtolength{\baselineskip}{-1.0pt}  {\sc Fig.~2} --- The rest-frame SED of SMM\,J2135$-$0102, a
typical high-redshift SMG at $z=2.3259$, showing the bright, narrow molecular and atomic emission lines seen in the
far-infrared and submillimetre.  The continuum component of the SED is derived from fits to the {\it Herschel} and
ground-based photometry  (Swinbank et al.\ 2010; Ivison et al.\ 2010a) corrected for the emission line contributions. 
Added to this are the emission lines derived from the observations of Danielson et al.\ (2011), or for  lines
short-ward of 100$\mu$m, from the predictions from the PDR model used in that work.  The observed [C{\sc ii}] line in
this source contributes 0.27\% of the L$_{\rm FIR}$ and along with other strong lines may make a significant
contribution to broad-band fluxes at certain redshifts (to better illustrate this, the inset shows the same SED
around the dust peak, with a linear flux scaling).  We also indicate the observed-frame pass-bands for the {\it
Herschel} PACS (160$\mu$m) and SPIRE (250, 350, 500$\mu$m), SCUBA-2 (450, 850$\mu$m) and LABOCA (870$\mu$m) filters.

}

\section{Results}

As an illustration of the effects of line contamination on the  broad-band fluxes of high-redshift SMGs we redshift
our template SED and at each redshift we find the fluxes which would be observed in the {\it Herschel} 160-, 250-,
350- and 500-$\mu$m bands, and the 450- and 850-$\mu$m SCUBA-2 bands (the latter matching the LABOCA 870-$\mu$m
filter).  In Fig.~3 we plot the fractional increase in these fluxes as a function of redshift for the
SMM\,J2135$-$0102 SED, compared to a line-free SED.  This figure demonstrates that even for relatively weak-lined
SEDs there can be moderate variations in the expected fluxes, 5--10\%, for many pass-bands at $z\gs 1$.  These line
contributions are comparable to the absolute flux calibration uncertainty of typical submillimetre maps (e.g.\ Weiss
et al.\ 2009) and the residual errors in source fluxes after correcting for confusion effects in these low-resolution 
maps (so-called flux boosting, Coppin et al.\ 2006), although unlike these two sources of flux error the line
contamination is redshift dependent.

However, given that we already know of examples of high-redshift sources with  stronger lines than our template SED
(Fig.~1), there is the potential for even larger variations. As an example we re-normalise the atomic lines in our
template SED so that [C{\sc ii}]\,158$\mu$m corresponds to 1\% of the bolometric emission and recalculate their
contribution to the broad-band fluxes.  We find that for such strong-lined emitters the emission lines can boost the
broad-band fluxes by $\sim 20$--40\% or greater (as shown by the right-hand scale in Fig.~3).  Such variation in the
apparent flux of sources due to emission lines falling in the pass-band have two obvious consequences:   i) the
presence of lines in only a subset of pass-bands may result in unusual colours for some combinations of filters at
certain redshifts; ii) the visibility of strong-lined sources at particular redshifts is enhanced, resulting in their
over-representation in flux-limited samples.  

%
%
\centerline{\psfig{file=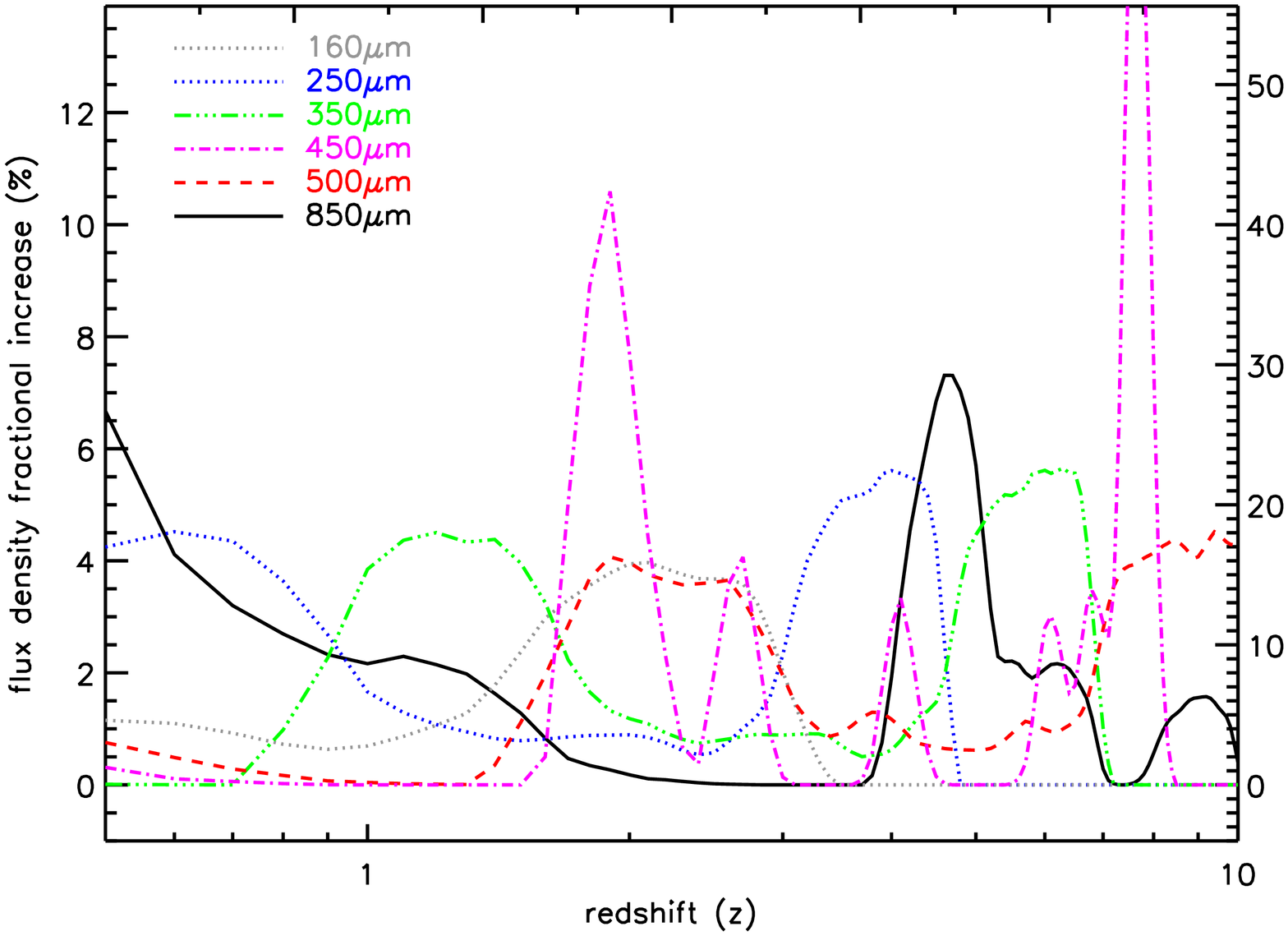,width=3.3in,angle=90}}
\noindent{\small\addtolength{\baselineskip}{-1.0pt}   Fig.~3 --- The contribution as a function of source redshift to
the continuum emission in the {\it Herschel} PACS and SPIRE, SCUBA-2 and LABOCA  bands, from the emission lines shown
in Fig.~2.  The influence of the [C{\sc ii}]\,158$\mu$m line can be seen in the 250-, 350-, 500- and 850-$\mu$m bands
at $z\sim $\,0.6, 1.2, 2.2 and 4.4 respectively with a contribution to the broad-band fluxes of 5--10\%.   Note that
these contributions are based on a  [C{\sc ii}] line comprising 0.27\% of the galaxy's L$_{\rm FIR}$.  The line
contributions will scale linearly with the line to far-infrared luminosity ratio, so sources with L$_{\rm
[CII]}/$\,L$_{\rm FIR}\gs 1$\% (Fig.~1) will have contributions $\gs 4\times$ larger, $\sim 20$--40\%, corresponding
to the right-hand flux scale.  We caution that the  contributions for the higher redshift sources in the shorter
wavelength filters (e.g.\ at $z\sim 4$ and $z\sim 6$ at 250 and 350$\mu$m respectively) are based on predicted,
rather than observed, line fluxes.
 
}
\smallskip

To show the influence of line contamination on properties derived from the broad-band photometry we look at the
effect of including emission lines on the dust temperature and luminosity estimates.  We fit the fluxes from our
redshifted template SED, with and without emission lines, with a modified black body with $\beta=1.5$ at the known
redshift and determine the characteristic temperature, T$_{\rm dust}$, and far-infrared luminosity, L$_{\rm FIR}$. 
Using the SMM\,J2135$-$0105 SED we find negligible effects on the derived temperatures and luminosities, $\ls 1$\,K
and $\ls 10$\% respectively, when complete, high-quality photometric data is available. However, sources with
stronger lines may suffer more significant biases, e.g.\ $\sim 30$\% in luminosity. Moreover, the increased flux in
the line-contaminated pass-bands, and hence the apparently higher significance of the detections in these pass-bands
coupled with incomplete photometric coverage, may  lead to even larger discrepancies.  Such effects would contribute
to the increased scatter in relationships such as the far-infrared--radio correlation at high redshifts (Ivison et
al.\ 2010b).   

We suspect that the most significant effect of the strong emission lines is likely to be on the apparent fluxes of
SMGs in those redshift ranges where a  line falls into the pass-band used for selection.   Seaquist et al.\ (2004)
showed this effect in local FIR-bright galaxies, where CO(3--2) emission typically makes a $\sim 25$\% contribution
to their 850-$\mu$m flux densities. Indeed, as their sample was {\it IRAS}-selected, this contribution is potentially
lower than it would have been in an 850-$\mu$m selected sample. To search for evidence of this effect at high
redshift we show in Fig.~4 the variation in apparent flux with redshift for a compilation from SCUBA or LABOCA
850-$\mu$m surveys.  We plot SMGs from the spectroscopically-identified  sample of Chapman et al.\ (2005); the SCUBA
map of GOODS-N with photometric redshifts from Pope et al.\ (2006); the LABOCA survey of the ECDFS (Weiss et al.\
2009) with photometric redshifts from Wardlow et al.\ (2011) and two cluster lens surveys (Smail et al.\ 2002;
Knudsen et al.\ 2008).   While these do not comprise a ``complete'' survey in any sense, they are representative of
our current knowledge of the flux--redshift distribution of SMGs selected at 850\,$\mu$m.  Binning these data up in
$\Delta z \sim 1$ redshift bins we see little variation in the median 850-$\mu$m flux of the SMGs with redshift out
to $z\sim 4$.  However, the final redshift bin, $z\sim $\,4--5, shows a broader distribution in submillimetre flux
with a slight offset to brighter fluxes.  We over-plot on this the expected variation in apparent flux with redshift,
arising from line contamination based on SEDs with L$_{\rm [CII]}/$\,L$_{\rm FIR}\sim $\,0.27, 1 and 5\%.  This
illustrates that a population of sources with high L$_{\rm [CII]}/$\,L$_{\rm FIR}$ ratios would result in a
measurable enhancement of the 850-$\mu$m broad-band fluxes of $z\sim 4$ SMGs.  Such behaviour might help explain the
suggestions that the brightest SMGs are typically at the highest redshifts (e.g.\ Ivison et al.\ 2002, 2007; Younger
et al.\ 2009;  but see Wardlow et al.\ 2011). However, we caution that there are a number of selection effects which
may be influencing this trend and that it requires direct confirmation through the  measurement of the [C{\sc ii}]
luminosities of these bright, high-redshift SMGs.
\medskip

%
%
\centerline{\psfig{file=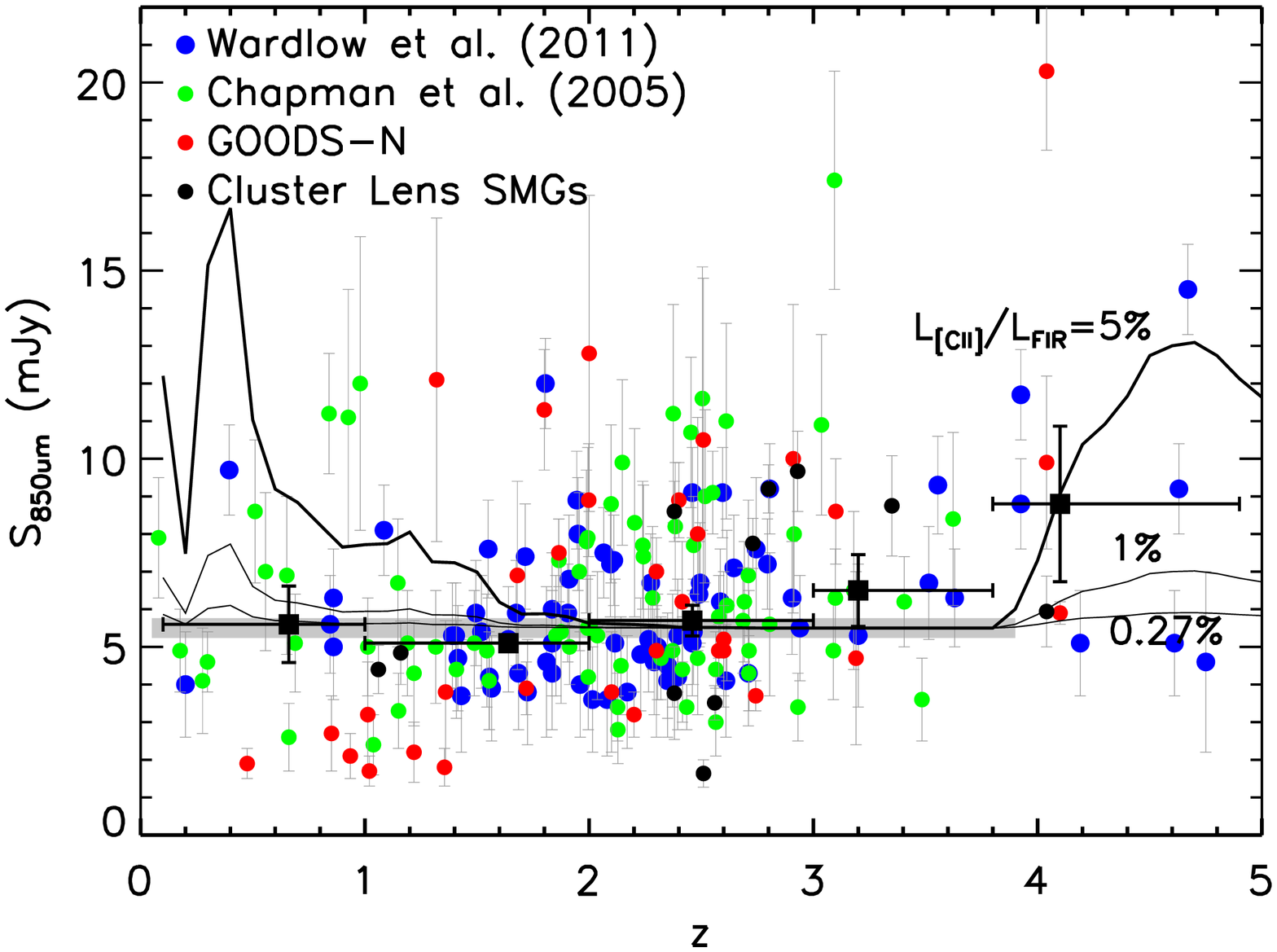,width=3.3in,angle=90}}
\noindent{\small\addtolength{\baselineskip}{-1.0pt}  {\sc Fig.~4} --- The variation in observed 850-$\mu$m flux with
redshift for a compilation of samples of SMGs from blank-field and cluster-lensing surveys using the SCUBA and LABOCA
cameras (Smail et al.\ 2002; Chapman et al.\ 2005; Pope et al.\ 2006; Knudsen et al.\ 2008; Wardlow et al.\ 2011). 
We plot the variation in median flux within $\Delta z\sim 1$ bins, the final bin from $z=$\,3.8--4.9 corresponds to
the redshifts where [C{\sc ii}]\,158$\mu$m falls in the 850\,$\mu$m pass-band.  Below $z\sim 4$ there is little
variation in the median flux as a function of redshift and we illustrate this by plotting the median flux of SMGs in
this redshift range and their 1-$\sigma$ dispersion as a grey bar.  In contrast, there is a suggestion that the
typical flux of SMGs at $z\gs 4$ may be higher, which may reflect the contribution of [C{\sc ii}] to their broad-band
fluxes. We over-plot as a solid line the expected variation with redshift in the median flux of an SMG with an SED
comparable to SMM\,J2135$-$0102, L$_{\rm [CII]}/$\,L$_{\rm FIR}\sim 0.27$\%,  as well as the tracks for SEDs with
[C{\sc ii}] line fluxes of L$_{\rm [CII]}/$\,L$_{\rm FIR}=$\,1 and 5\%. If a subset of $z\gs 4$ SMGs exhibit enhanced
[C{\sc ii}] emission then this could result in the apparent brightening at 850$\mu$m of SMGs at these redshifts.

}
\medskip

To determine the potential extent of an emission-line bias in submillimetre surveys we construct a toy evolutionary
model and use this to determine how the redshift distribution of SMGs changes when we include emission lines in the
SEDs used.  Our evolutionary model is the best-fitting model from Blain et al.\ (1999), ``Model Peak-G'',  which
comprises strong Gaussian luminosity evolution characterised by a broad peak at $z\sim 2$ with $\sigma_z \sim 1$. We
apply this evolution to a population of sources with a single SED and a luminosity function matching that measured
locally using {\it IRAS} by Saunders et al.\ (1990).  We then tune the parameters of the model to reproduce the
850-$\mu$m number counts and the broad properties of the redshift distribution of SMGs from Wardlow et al.\ (2011). 
We stress that this model is not physically well-motivated or robust, rather it is intended to be a simple parametric
representation of the behaviour of the SMG population. 

%
%
\centerline{\psfig{file=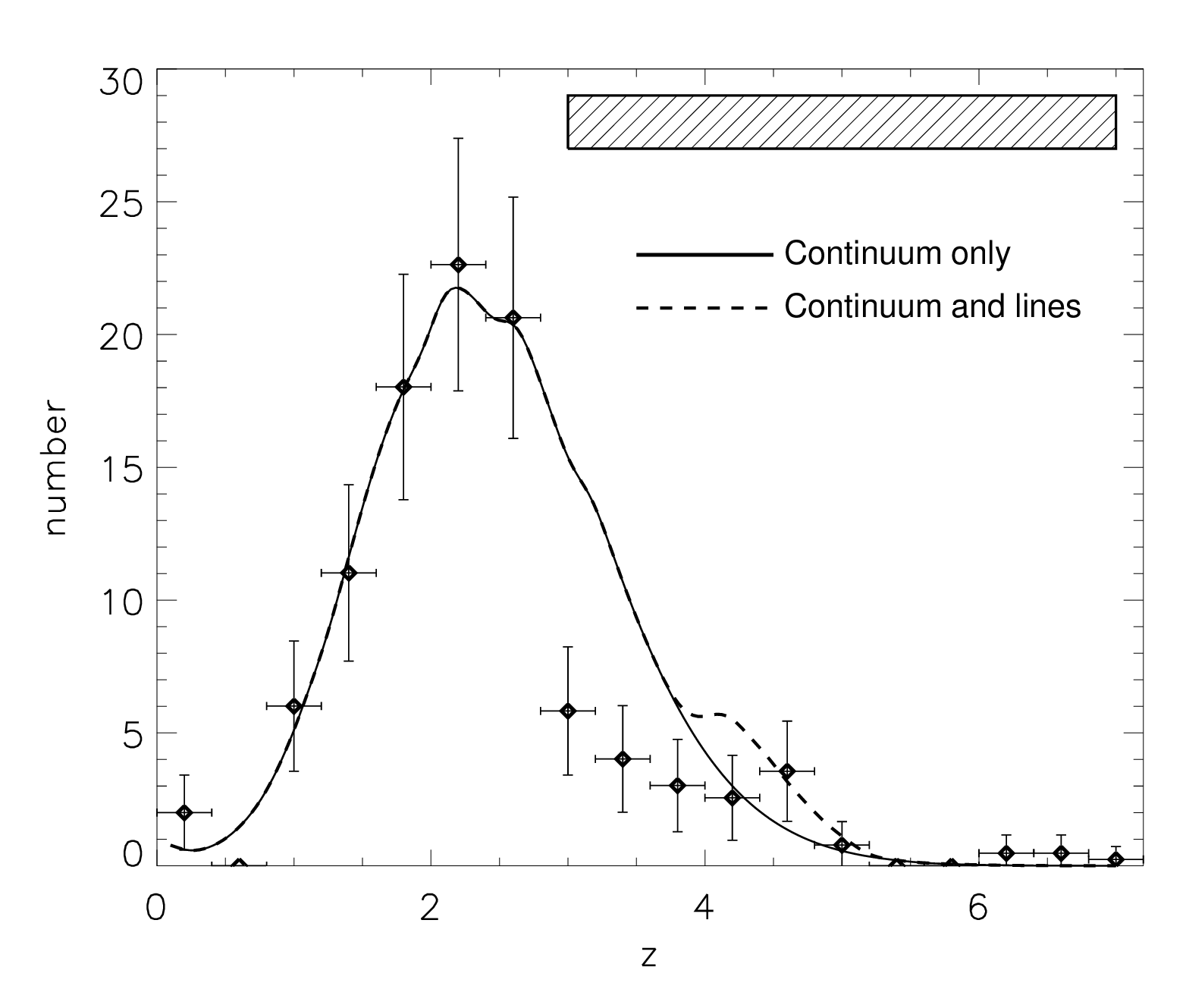,width=3.3in,angle=0}}
\noindent{\small\addtolength{\baselineskip}{-1.0pt}  Fig.~5 --- The predictions of a toy evolutionary model for the
redshift distribution of an 850-$\mu$m-selected sample with S$_{850\mu\rm m}\geq 5$\,mJy.  The model is based on
those from Blain et al.\ (1999) and it is tuned to reproduce the 850-$\mu$m counts and the broad properties of the
redshift distribution for 850-$\mu$m sources from Wardlow et al.\ (2011), which is plotted here as binned points (the
shaded box illustrates the extent and likely redshift distribution of the incompleteness in the Wardlow survey).  We
plot two versions of the model, using the continuum SED of SMM\,J2135$-$0102, with  and without  emission lines.  As
can be seen the two models deviate at $z\sim $\,4--5, due to the presence of strong [C{\sc ii}]\,158$\mu$m in the
850-$\mu$m pass-band, which roughly doubles the  numbers of sources detected at these redshifts by increasing the
measured 850-$\mu$m fluxes of sources below the nominal continuum flux limit. 
 
}
\medskip

We show in Fig.~5 the predicted redshift distributions for an 850-$\mu$m selected sample of sources with
S$_{850\mu\rm m}\geq 5$\,mJy, assuming the SED of SMM\,J2135$-$0102 with and without emission lines.  As can be seen
the two models deviate at $z\sim $\,4--5, due to the presence of strong [C{\sc ii}]\,158$\mu$m in the 850-$\mu$m
pass-band. The number density of  850-$\mu$m selected sources at $z\sim 4$--5 increases by a factor of $\sim 2\times$
when we include the  emission lines in the SED.  This excess arises due to the steep slope of the far-infrared
luminosity function at the relevant flux levels and the modest boost provided by the line emission to the measured
broad-band fluxes. The factor of $\sim 2\times$ excess is representative of that expected in 850-$\mu$m samples if
the L$_{\rm [CII]}/$\,L$_{\rm FIR}$ ratio for SMM\,J2135$-$0102 is typical of high-redshift SMGs, however, larger
enhancements are  possible if the population has a systematically higher  L$_{\rm [CII]}/$\,L$_{\rm FIR}$ ratio.

\section{Discussion and Conclusions}

We have demonstrated that the strong far-infrared emission lines being uncovered in high-redshift sources have a
potential impact on the broad-band submillimetre observations of these populations.   As we lack a complete census of
the emission-line properties of high-redshift SMGs we have chosen to adopt the well-studied SMM\,J2135$-$0102 as our
template SED.  However, we note that based on local galaxy populations, it is possible that there are galaxies with
significantly stronger line emission than this system.  Such strong line emitters will be preferentially selected by
flux-limited continuum surveys at specific redshifts.  For example, 850-$\mu$m-selected samples of SMGs at $z=$\,4--5
may include an over-abundance of sources with high  L$_{\rm [CII]}/$\,L$_{\rm FIR}$ ratios (Fig.~4), while SPIRE
surveys may show enhanced source numbers at $z\sim $\,0.6, 1.2 or 2.1 for 250-, 350- and 500-$\mu$m selected samples.  

The influence of emission lines may also lead to difficulties in identifying the counterparts to the 
highest-redshift SMGs from 850-$\mu$m surveys with submillimetre interferometers. Due to the presence of the strong
[C{\sc ii}]\,158$\mu$m line, the apparent continuum flux of these SMGs may be overestimated, and  unless the line
falls in the relatively narrow frequency coverage of the interferometer there is a strong chance that the source may
not be detected. 

In the near future high-frequency ALMA surveys will start to detect [C{\sc ii}] emission in sources from $z\gs 1$ out
to high redshifts (Walter \& Carilli 2008). These surveys will have sufficient sensitivity to detect Lyman-break
galaxies with star-formation rates of $\ls 10$\,M$_\odot$\,yr$^{-1}$ and metallicities below $0.1Z_\odot$ (Mannucci
et al.\ 2009).  If  the same enhancement of  L$_{\rm [CII]}/$\,L$_{\rm FIR}$ seen in low-metallicity galaxies locally
(e.g.\ Madden et al.\ 1997; Rubin et al.\ 2009) is also seen in these younger, lower luminosity and lower metallicity
galaxies at high redshift,  then line contamination may be an issue for these surveys.  However, the spectral
resolution of ALMA should allow such strong lines to be  identified and their contribution to the broad-band flux
assessed.

In summary, there is growing observational evidence suggesting that high-redshift SMGs exhibit strong atomic cooling
lines of C, N and O.  The luminosities of these lines can potentially exceed 1\% of the total bolometric emission
from these luminous galaxies.  We show that such strong lines will have a marked influence on the broad-band flux
densities of SMGs at specific redshifts, depending upon the choice of broad-band filter.  We find  maximum  line
contamination contributions of  $\sim 30X$\% to the flux densities in the {\it Herschel} PACS and SPIRE and SCUBA-2
bands for a line contributing $X$\% of the bolometric luminosity.  For the strongest line emitters this contribution
to their broad-band fluxes will exceed the uncertainties due to residual errors from flux de-boosting corrections
(arising from source confusion) or absolute calibration errors.  Unlike these other factors the line contamination
has a redshift dependency and as we show, by comparing the flux--redshift and redshift distributions for  850-$\mu$m 
surveys to simple models with and without emission lines, the presence of emission lines will lead to an excess of
bright sources at $z=4$--5, which can be observationally tested with forthcoming surveys.  Of course the most direct
route to test the importance of line emission in continuum surveys for SMGs is to measure the strength of the lines
falling in the selection pass-band for a particular source.  The broad frequency coverage and high sensitivity of
ALMA mean that it should be easy to do so in future.  If strong line-emitting sources do exist then narrow-band
surveys in the submillimetre waveband may be an efficient route to survey for them using bespoke narrow-band filters
or Fourier Transform Spectrometers such as FTS-2 on SCUBA-2.

\section*{Acknowledgements}
We thank the referee for a constructive report which improved the content and presentation of this paper.  We also
thank Scott Chapman, Alastair Edge, Jim Geach, Dieter Lutz, Rowin Meijerink, Padelis Papadopoulos and Paul van der
Werf for useful discussions. IRS, RJI and EI acknowledge support from STFC, AMS acknowledges support through an STFC
Advanced Fellowship.

\end{document}